\documentclass[useAMS,usenatbib,twocolumn]{mnras}
\usepackage{graphicx,url,amssymb,amsmath,color,units,wasysym,epsfig,epstopdf,enumerate,tabularx, subfigure,hyperref}

 \usepackage{float}
\usepackage{newtxtext,newtxmath}
\usepackage[T1]{fontenc}
\usepackage{ae,aecompl}
\usepackage{appendix}
\usepackage[dvipsnames]{xcolor}
\usepackage[normalem]{ulem}
\usepackage[bottom]{footmisc}

\begin{document}

\title[Pulsar timing with dynamic fitting]{Improving pulsar-timing solutions through dynamic pulse fitting}

\author[]{
\parbox{\linewidth}{Rowina S. Nathan$^{1,2}$, Matthew T. Miles$^{2,3}$, Gregory Ashton$^{4}$, Paul D. Lasky$^{1, 2}$, Eric Thrane$^{1,2}$,\\
Daniel J. Reardon$^{2,3}$, Ryan M. Shannon$^{2,3}$, Andrew D. Cameron$^{2,3}$}\\
$^{1}$School of Physics and Astronomy, Monash University, VIC 3800, Australia\\
$^{2}$The ARC Center of Excellence for Gravitational Wave Discovery – OzGrav, Hawthorn VIC 3122, Australia\\
$^{3}$Centre for Astrophysics and Supercomputing, Swinburne University of Technology, PO Box 218, Hawthorn, VIC 3122, Australia\\
$^{4}$Department of Physics, Royal Holloway, University of London, TW20 0EX, United Kingdom
}

\maketitle

\begin{abstract}
Precision pulsar timing is integral to the detection of the nanohertz stochastic gravitational-wave background as well as understanding the physics of neutron stars. Conventional pulsar timing often uses fixed time and frequency-averaged templates to determine the pulse times of arrival, which can lead to reduced accuracy when the pulse profile evolves over time. We illustrate a dynamic timing method that fits each observing epoch using basis functions. By fitting each epoch separately, we allow for the evolution of the pulse shape epoch to epoch. We apply our method to PSR J1103$-$5403 and find evidence that it undergoes mode changing, making it the fourth millisecond pulsar to exhibit such behaviour. 
Our method, which is able to identify and time a single mode, yields a timing solution with a root-mean-square error of $\unit[1.343]{\mu s}$, a factor of 1.78 improvement over template fitting on both modes. In addition, the white-noise amplitude is reduced 4.3 times, suggesting that fitting the full data set causes the mode changing to be incorrectly classified as white noise. This reduction in white noise boosts the signal-to-noise ratio of a gravitational-wave background signal for this particular pulsar by 32\%.
We discuss the possible applications for this method of timing to study pulsar magnetospheres and further improve the sensitivity of searches for nanohertz gravitational waves.
\end{abstract}


\section{Introduction}

Due to their remarkable regularity, pulsars are exceptionally accurate clocks.
Pulsar timing is therefore useful across many areas of physics. Timing many stable pulsars across the sky in a pulsar timing array can detect low-frequency nanohertz gravitational waves \citep{Hellingsanddowns} from sources such as supermassive binary black holes \citep{Sesana2004, Kocsis2011, taylor2017, Burke2019} and phase transitions in the early Universe \citep[e.g.,][]{Starobinsky1980, Grishchuk2005, Lasky2016}. 
The most recent data sets from three of the worlds major pulsar-timing arrays, the European Pulsar Timing Array  \citep[EPTA;][]{EPTA}, the North American Nanohertz Observatory for Gravitational Waves \citep[NANOGrav;][]{NANOGrav}, and the Parkes Pulsar Timing Array \citep[PPTA;][]{PPTA} have shown evidence of a common, red noise process in the residuals of their pulsar arrival times \citep{EPTACRN,PPTACRN, NanoGravCRN}. Common red noise was also found when this data was combined by the International Pulsar Timing Array \citep[IPTA;][]{IPTACRN}. Common red noise can arise due to the stochastic gravitational-wave background.
However, in order to make an unambiguous detection, one must observe an angular correlation function consistent with the Hellings and Downs curve \citep{Hellingsanddowns}, a quadrupolar correlation between timing delays across the sky.  With new data releases imminent, the detection of nanohertz gravitational waves may be around the corner. 

Current pulsar-timing methods employ a single matched-filter template to calculate pulse times of arrival\footnote{Frequency-dependant templates are sometime used, either with multiple templates across different subbands \citep[e.g.,][]{vanStraten2006, Liu2014} or by fitting functional forms to the template across the frequency band \citep[e.g.,][]{Pennucci2014, Pennucci2019}.}
Often, the template is created by averaging together many of the observed pulses.
This single, static template is then used to time all the pulses.
However, all pulsars exhibit at least some degree of pulse-to-pulse shape variation \citep{Parthasarathy2021jitter}, which conventional pulsar timing methods are not able to account for. 

There are a number of phenomena that are known to cause obvious changes in pulse shape.  For example, giant pulses are extraordinarily bright and narrow outbursts of radio flux \citep{Staelin1968GPs, Geyer2021GiantP, Caleb2022GiantP}. Pulses can be temporarily broadened or lensed by the interstellar medium \citep{Rickett_ISM, Shannon2017ISM, Bilous2019broad, Lin2021broad, Askew2022ISM}. Geodetic precession, where the strong gravitational field of the pulsar results in spin-orbit coupling, causes pulse shapes to evolve over time \citep{Kramer1998GeoP, Fonseca2014GeoP, Desvignes2019GeoP, Venkatraman2019GeoP, Noutos2020GeoP}. A pulsar can undergo a period of mode-changing, in which it switches between two or more characteristic pulse shapes \citep{Bartel1982Mode, Wang2007Mode, Miles2022Mode}. On occasion, pulses have been seen to cease completely, in a process known as nulling \citep{Backer1970Null, Gajjar2012Null}.

There are also more subtle pulse-shape changes, such as stochastic changes known as ``jitter'' \citep[][]{Shannon2014jitter, Parthasarathy2021jitter} and slow evolution in pulse shape due to sub-pulse drifting \citep{Drake_subpulse, backer_subpulse, Chen_subpulse}.
A single pulse profile template does not capture the pulse-shape variation from these and other effects. Since the average template may not be a good match for some pulses, the estimated pulse time of arrival can be significantly wrong, leading to increased errors in the timing solution. Pulse-shape variation therefore is a source of noise for pulsar-timing arrays, reducing our ability to detect nanohertz gravitational waves.

While our primary goal is to reduce the timing solution residuals for gravitational-wave searches, modelling the evolution of pulse shape is interesting in its own right. By studying how pulse shape varies over time, it may be possible to gain a better understanding of the pulsar magnetosphere and/or pulsar emission \citep{Rankin1986shapechange, Cairns2004shapechange, Janagal2022shapechange}. Moreover, pulsar timing allows for tests of general relativity in the strong field limit \citep{Kramer2006, Freire2012, Archibald2018, Voisin2020, KramerGR,  Kramer2021} and it provides us with an improved understanding of the neutron star equation of state \citep{Demorest2010, Antioniadis2013EOS, Miller2021EOS, Riley2021EOS, Fonseca2021EOS} and thus the behaviour of matter at extreme densities \citep{Oppenheimer1939, Kurkela2014, Ozel2016}.

All pulsars exhibit pulse-shape variations, however large shape-change events are less common in millisecond pulsars; they are mostly stable which makes them good candidates for long-term pulsar timing. There are currently three millisecond pulsars known to mode-change: PSR B1957$+$20 \citep{Mahajan20181957}, PSR J0621$+$1002 \citep{Wang20210621} and PSR J1909$-$3744 \citep{Miles2022Mode}.
PSR J1103$-$5403 is a pulsar observed by the MeerKAT telescope as part of the MeerKAT Pulsar Timing Array \citep{MeerKAT}. We show that this pulsar exhibits the characteristics of a mode changing pulsar, as it has a group of early arriving outliers in the timing solution residuals. This millisecond pulsar has a period of only $\sim\unit[3.4]{ms}$ and is a good candidate for a timing array pulsar.
However, the mode changing severely restricts its timing accuracy.
This makes it an ideal test case for an alternate timing method that is better able to constrain pulse-shape variability.

\cite{Lentati2015} developed a profile-domain timing method where individual time-averaged epochs were fit, allowing for simultaneous estimation of the timing model, dispersion measure variations and pulse-profile evolution. When this was implemented with broad-band frequency evolution, an improvement of up to 40\% was seen in pulsar timing precision \citep{Lentati2017}.
Pulse-to-pulse timing has already been shown to measure the glitch rise time of the Vela Pulsar \citep{vela_glitch}.
However, pulse-to-pulse observations are only available for the brightest millisecond pulsars. It is therefore common to use epochs for timing, where an observing period is folded and summed over many pulse periods to increase the brightness of the pulse.
We implement here an epoch-to-epoch fitting method, in order to determine if the flexibility of this method provides insights on time-averaged data.

In this paper, we present a pulsar-timing method that allows for and is sensitive to pulse-shape variation, using PSR J1103$-$5403 as a case study. We are able to confidently determine outliers in pulse shape, the removal of which from the total data results in a reduction of the timing solution root-mean-square error (RMS; this error arises from the difference between pulse times of arrival predicted by the timing solution and observations). We reduce the RMS of this pulsar by a factor 1.78 consequently improving the sensitivity of PSR J1103$-$5403 to the gravitational-wave background by 32\%.
We describe how our method can be more broadly applied to other pulsars. 
The remainder of this paper is organized as follows.
In Section \ref{fitting}, we describe our mathematical formalism and individual epoch fitting. In Section \ref{timingsolution}, we present the results of our analysis of PSR J1103$-$5403 and compare them to the matched-filter template method. We then use the parameters fit to each pulse to characterise the shapes present in the mode changing, and produce a single mode timing solution. Finally, in Section \ref{discussion} we discuss the implications of our results and consider avenues for future work. 

\section{Individual epoch fitting}\label{fitting}

In order to fit the flux profile of each epoch, we fit a sum of basis functions \citep[similar to][]{Kramer1994, Lentati2015, Padmanabh2021, Cameron2022}. The parameters of these basis functions are fit independently for each epoch. 
We employ a Bayesian approach, using nested sampling \citep{Skilling2004Nested, Skilling2009Nested} to explore the parameter space.

\begin{figure}
	\centering 
	\includegraphics[width=\linewidth]{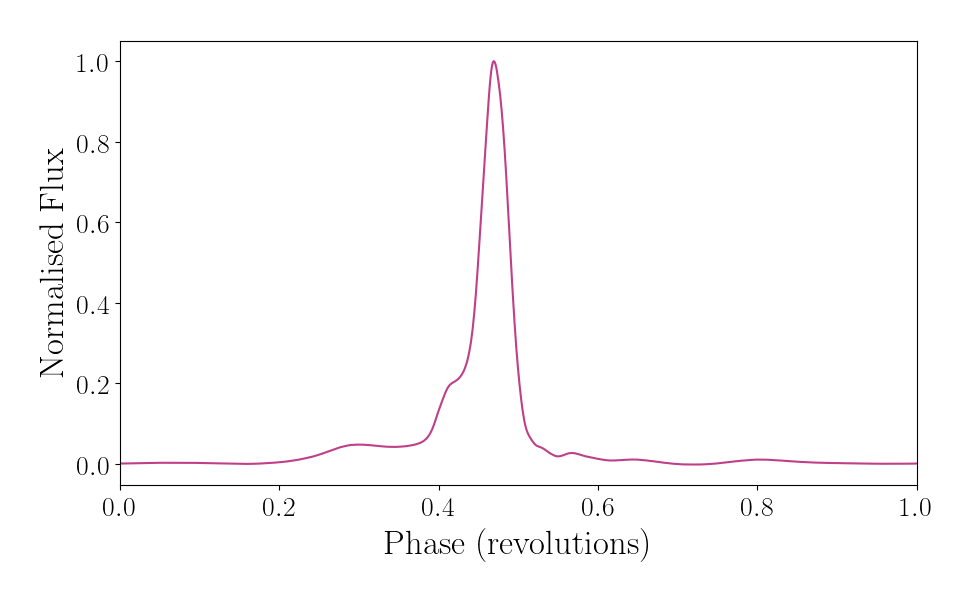}
	\caption{The smoothed total intensity profile of PSR J1103$-$5403. This is the average of 52 observing epochs from across three years, which is then smoothed to emphasise the main pulse features. }
	\label{fig:smoothpulse}
\end{figure} 

In Fig.~\ref{fig:smoothpulse} we show the averaged pulse profile for PSR J1103$-$5403, which has been smoothed to reduce noise using \texttt{psrsmooth} \citep{Demorest2013}. By visually inspecting the pulse shape, we identify three main pulse features: the shoulder at $\sim 0.30$ in phase, the leading edge of the pulse at $\sim0.42$ and the brightest portion of the pulse at $\sim0.48$\footnote{Following conventions in pulsar timing literature, we measure phase in units of revolutions (from zero to one).}.
Based on this morphology, we choose a sum of three Gaussians as the basis functions to model the pulse profile,
\begin{equation}
    \label{eqn:gaussian}
    F(\phi)=\sum_{i=0}^{2}C_ie^{-(\phi-\phi_i)^2/\beta_i^2}+\sum_{j=0}^{2}B_j(\phi-0.5)^j ,
\end{equation}
where $F(\phi)$ is the flux at phase $\phi$, $\phi_i$ is the phase-centering of each Gaussian, $C_i$ is the amplitude and $\beta_i$ is the width. The second summation is included to model the baseline flux as a sum of polynomials of order $j$. The scale of the baseline flux for each polynomial is modeled by $B_j$ with a subtraction of 0.5 in phase to ensure centering in the middle of the pulse.

We use Fig.~\ref{fig:smoothpulse} to inform our priors, summarised in Tab.~\ref{tab:priors}. A priori we know that the width of the Gaussians should be small compared to the pulse period motivating priors that are uniform in $\log_{10}(\beta_i)$. We place a minimum on each $\beta_i$ that is five times the width of the time bins to prevent the model from fitting small noise fluctuations. The maximum on $\beta_i$ prevents the Gaussian fitting the shoulder of the pulse (Gaussian 1) from interfering with the main pulse (which we use to determine the time of arrival as discussed below). The prior on both the Gaussian means $\phi_i$ and amplitudes $C_i$ is a bounded uniform prior, forcing $C_0<C_1<C_2$ and $\phi_0<\phi_1<\phi_2$. This prior on the $\{\phi_i\}$ ensures that the Gaussians do not have multimodal posteriors due to swapping order. The constraint on the amplitudes are motivated by the average pulse profile and ensures the model does not fit noise features to the right of the main pulse. 

We also explore other potential models. 
For example, we fit a model with two Gaussians, but find the Bayesian evidence prefer a three-component model. 
We fit a four-Gaussian model and find that the additional Gaussian increases variation in the phase-centering of the largest-amplitude Gaussian, increasing uncertainty in the timing solution.
We attempt to model the epoch profile with a single higher order shapelet, as \citet{Lentati2017} \citep[see ][for a definition of shapelets]{refregier_shapelets} or sums of multiple higher order shapelets, but again find for these models that the data prefers the three Gaussian model, or that the computing expense is too great. As the pulse profile differs greatly between pulsars, this style of investigation is likely required for dynamic pulse fitting on all pulsars.

The fitting is done with \texttt{Kookaburra} \citep{KookaburraCode}, an open-source python library. This library allows for easy definition of the priors and basis functions, and uses \texttt{Bilby} \citep{bilby2019} for the fits themselves. 
\texttt{Kookaburra} fits pulses using shapelets (Gaussians are zero-th order shapelets), but users may define other basis functions as well. \texttt{Kookaburra} outputs the posterior of the parameters, as well as a Bayes factor comparing the signal hypothesis (that the pulses consists of linear combination of basis functions) with the noise hypothesis that no pulse is present.

We fit our model to the data using nested sampling. \texttt{Kookaburra} and \texttt{Bilby} allow for the use of a number of sampling algorithms; we use the nested sampling algorithm \texttt{pymultinest} \citep{multinest, pymultinest}. We individually fit 52 de-dispersed, frequency and polarisation averaged, time-folded observing epochs from PSR J1103$-$5403. We use observations taken using the MeerKAT radio telescope L-band receiver, collecting between 856 and 1712 MHz \citep{meerkat_band}. The observations have a nominal cadence of two weeks \citep{MeerKAT}, from 2$^\text{nd}$ August 2019 to 25$^\text{th}$ September 2021. 

\begin{figure}
	\centering 
	\includegraphics[width=\linewidth]{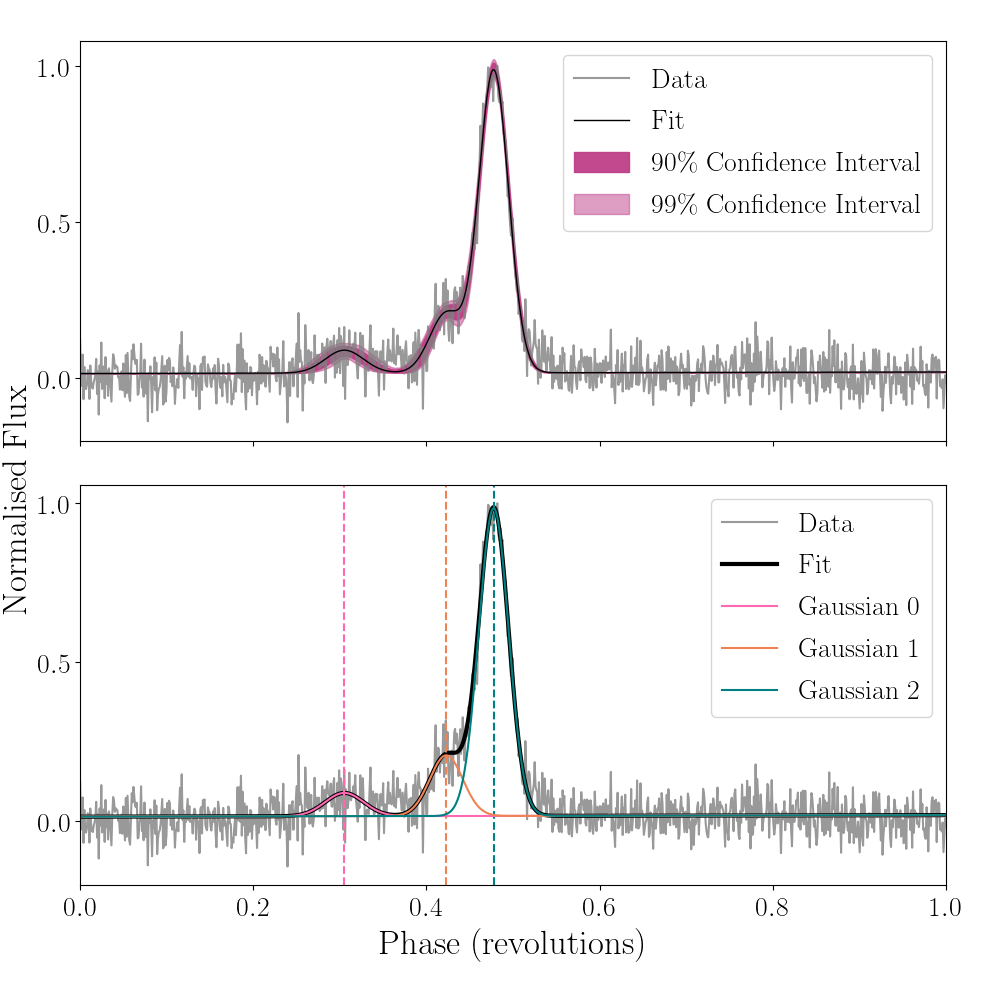}
	\caption{An example of one of the 52 individual pulse fits (this is the observing epoch from 12$^\text{th}$ March 2020). The top panel shows the maximum-likelihood fit in black, with $90\%$ and $99\%$ confidence intervals from the posterior distribution of the fit shown in pink. The bottom panel shows the three Gaussians (shown in orange, teal and pink) summed for the maximum likelihood fit (black). The pulse data is shown in grey. }
	\label{fig:examplepulsefit}
\end{figure}

\begin{figure}
	\centering 
	\includegraphics[width=\linewidth]{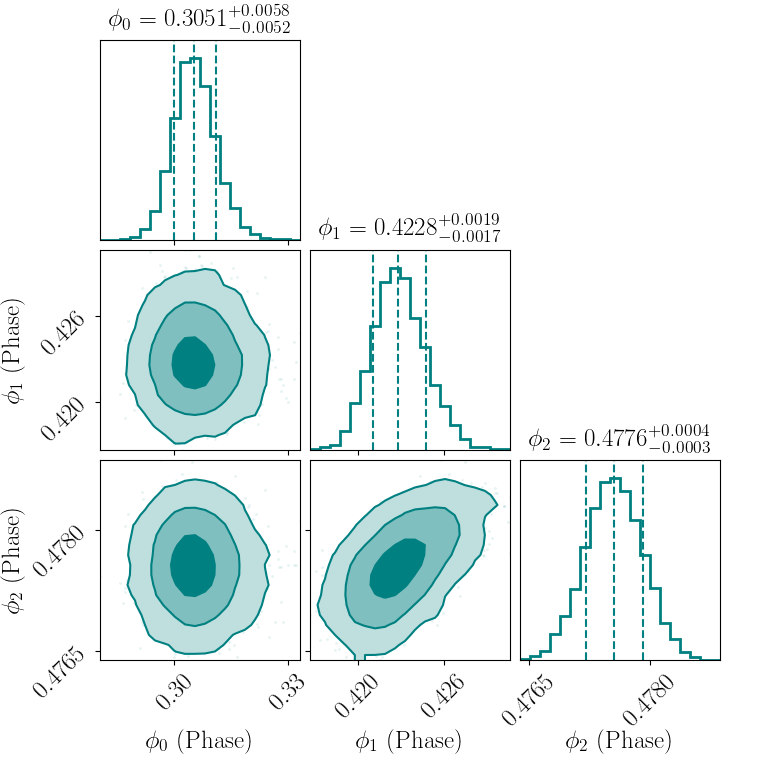}
	\caption{Posterior distributions for the centers of each Gaussian $\phi_i$ fit to the pulse in Fig.~\ref{fig:examplepulsefit}.}
	\label{fig:cornertoa}
\end{figure}

We show a fit to an example epoch (observed on 12$^\text{th}$ March 2020) in Fig.~\ref{fig:examplepulsefit}. In both panels the black curve shows the maximum-likelihood model fit and the grey curve shows the flux data. The top panel shows the $90\%$ and $99\%$ confidence intervals from the posterior distribution of the sampling in pink and the bottom panel shows the three Gaussians making up the maximum-likelihood fit. The pink Gaussian characterises the shoulder, the orange characterises the leading edge of the main pulse, and the teal Gaussian fits the brightest portion of the pulse. Fig.~\ref{fig:cornertoa} shows the posterior distribution for the mean of each Gaussian\footnote{The corresponding corner plot showing the posteriors for all parameters is shown in Appendix \ref{AppendixB} in Fig.~\ref{fig:corner}.}.

After fitting for the pulse profile in individual epochs, we construct a timing solution for PSR J1103$-$5403. 
There are numerous ways one could define the time of arrival. 
We choose to identify the time of arrival as the maximum-likelihood estimate for $\phi_2$, the mean of the third and most prominent Gaussian peak (denoted by the dashed teal line in Fig.~\ref{fig:examplepulsefit}). 
Our rationale is that this prominent feature of the pulse profile is likely the most stable over long periods of time.
We convert $\phi_2$, which is measured in phase, into a time of arrival 
in Modified Julian Day (MJD) in order to fit a timing solution. 
We record the one-sigma credible interval for $\phi_2$; this is an estimate for our uncertainty on the pulse arrival time\footnote{We use a single symmetric error estimate as \texttt{Tempo2} does not take a full posterior.}. 
The list of times of arrival (and their associated uncertainties) are passed to \texttt{tempo2} \citep{Edwards2006tempo, tempo2_chi}, which finds the best-fit pulsar model. This is achieved by a chi-squared minimisation of the residuals \citep[see][]{tempo2_chi}. 

\section{Timing solutions}\label{timingsolution}
\begin{figure}
	\centering 
	\includegraphics[width=\linewidth]{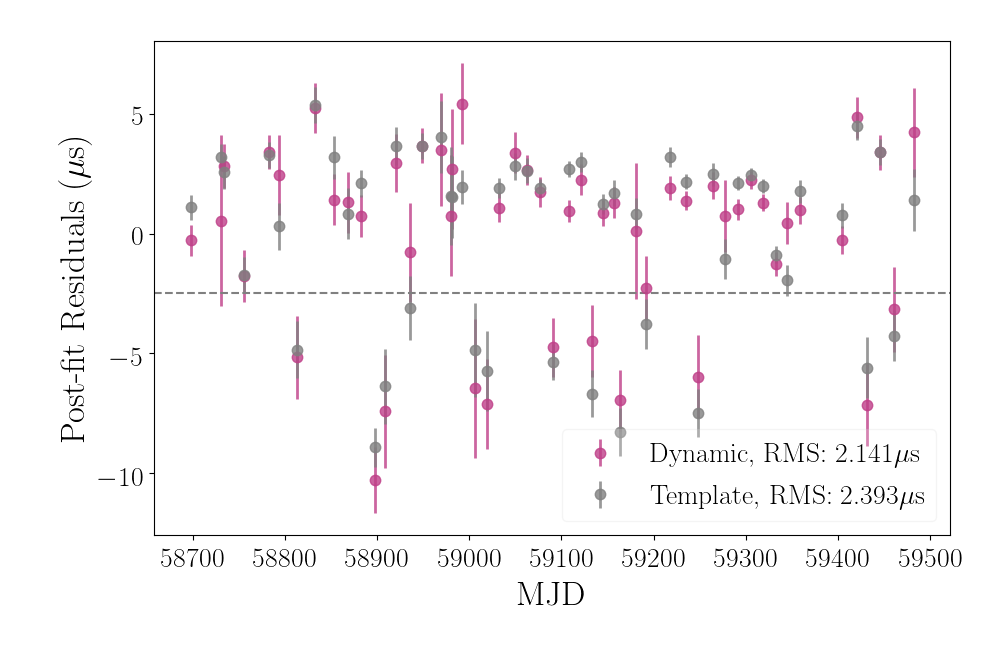}
	\caption{A comparison of the timing solution produced by template fitting (grey) and dynamic basis-function fitting with \texttt{Kookaburra} (pink). Our method is flexible to changes in pulse shape while maintaining timing accuracy. Our timing solution improves on matched-filter template methods, with a smaller root mean squared error of $\unit[2.141]{\mu s}$. Both timing solutions show evidence of mode changing, highlighted by the grey line dividing the two groups of residuals.}
	\label{fig:kookpsr}
\end{figure}

We show our timing solution in Fig.~\ref{fig:kookpsr}. The pink points represent the post-fit residuals obtained with our three-Gaussian fit. We compare our dynamic method to matched-filter template times of arrival obtained by \texttt{PSRchive} \citep{psrchive}. We fit the template using the default \texttt{PSRchive} fitting method \citep[Fourier Phase Gradient, see][]{taylor_PGS}. This fit is represented by the grey times of arrival in Fig.~\ref{fig:kookpsr}.
The two methods yield qualitatively similar timing solutions.
However, our three-Gaussian fit yields a somewhat lower RMS: $\unit[2.141]{\mu s}$ down from $\unit[2.393]{\mu s}$.
We attribute this reduction in RMS to the flexibility of our pulse profile fits, which we posit yield a more accurate fit on average.

The full power of our pipeline is yet to be demonstrated as additional improvement in the RMS is possible using analysis of the pulse profile shape.
In Fig.~\ref{fig:kookpsr}, we see that there is evidence of mode changing in PSR J1103$-$5403, which is evident as a tendency for the post-fit residuals to cluster around two distinct values: one at zero and one at $\approx\unit[-7.5]{\mu s}$. The dashed line on Fig.~\ref{fig:kookpsr} differentiates between the two modes.

\begin{figure}
	\centering 
	\includegraphics[width=\linewidth]{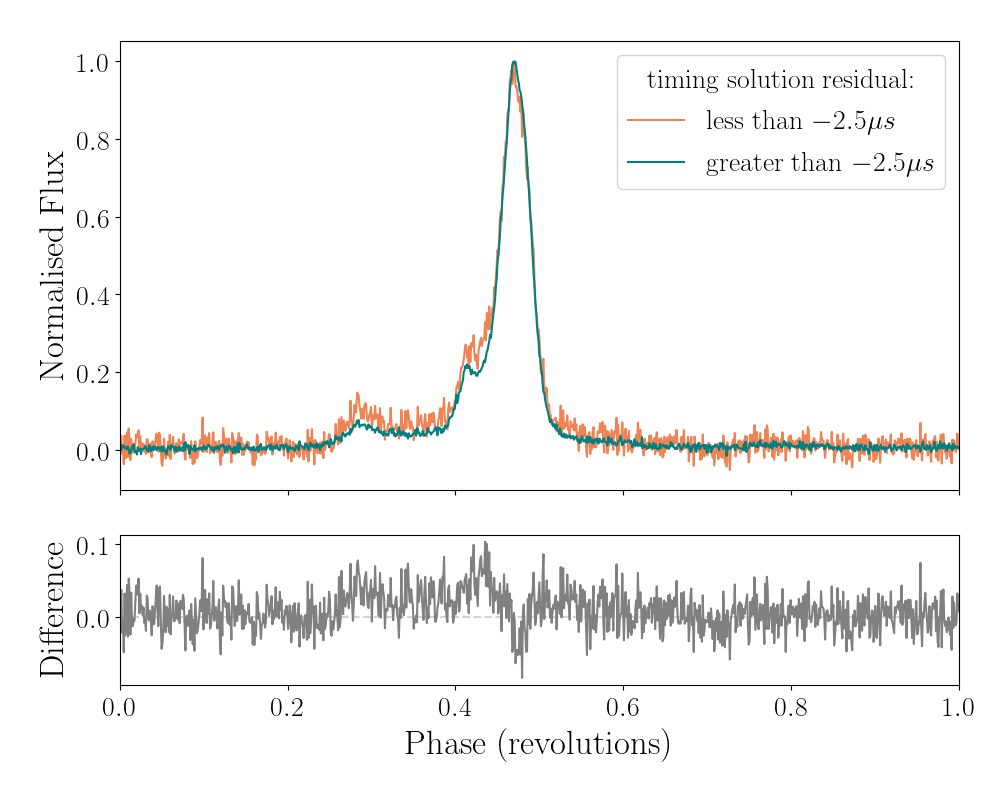}
	\caption{The average profile shapes of the two proposed modes, separated by the dashed line in Fig~.\ref{fig:kookpsr}. The top panel contains a teal curve comprised of epochs with a timing solution residual greater than $\unit[-2.5]{\mu s}$ (Mode A), whereas the orange curve is epochs with timing solution residuals less than $\unit[-2.5]{\mu s}$ (Mode B). The bottom panel shows the difference between these two mode in grey. There are noticeable differences in the shapes of these profiles at the leading edge ($\sim 0.42$ in phase) and the shoulder ($\sim 0.30$ in phase).}
	\label{fig:modeprofile}
\end{figure}

\begin{figure}
	\centering 
	\includegraphics[width=\linewidth]{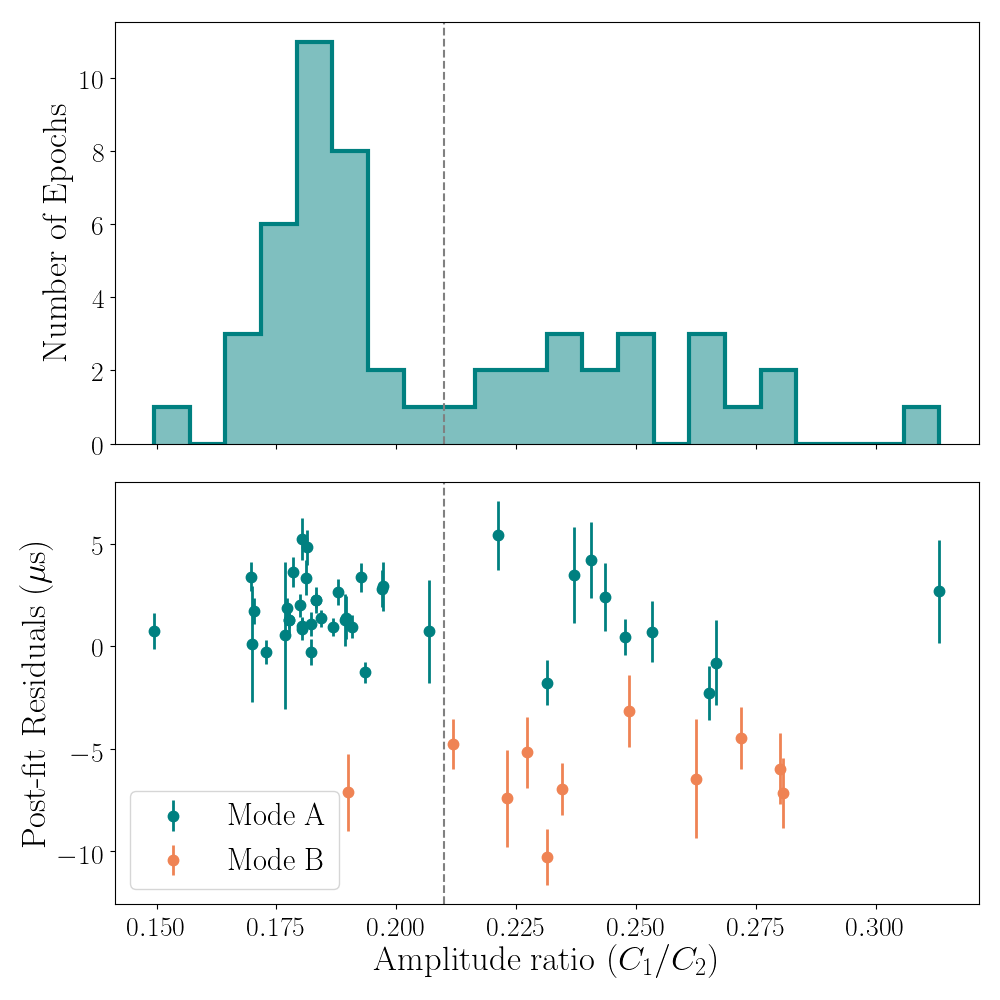}
	\caption{The ratio of Gaussian-peak-height one to Gaussian-peak-height two $C_1/C_2$. The top panel shows the bimodality arising from the different pulse shapes of Mode A and Mode B, which we use to design our cut. The bottom panel shows the timing solution residuals versus $C_1/C_2$, highlighting the effect of the cut to obtain a more pure sample of Mode A pulses.}
	\label{fig:modechange}
\end{figure}

To investigate the shape of the two modes, we plot the average profile of the two groups in Fig.~\ref{fig:modeprofile}. The difference in shape exhibited in Fig.~\ref{fig:modeprofile} suggests that the Gaussian fit to the leading edge of the pulse (Gaussian 1) may be able to differentiate between the modes. To further examine this, in the top panel of Fig.~\ref{fig:modechange} we show a histogram of $C_1/C_2$, the ratio of the height of the first Gaussian $C_1$ to the height of the second Gaussian $C_2$. The total flux varies between epochs, thus this ratio allows comparison between epochs.
This bimodal plot suggests there could be two distinct sub-populations of pulse profiles, delineated by the vertical grey dashed line. 
Plotting $C_1/C_2$ versus residual in  the bottom panel of Fig.~\ref{fig:modechange}, we see that the different sub-populations hinted at in Fig.~\ref{fig:kookpsr} and Fig.~\ref{fig:modeprofile} are strongly correlated with the different pulse profile modes \citep[similar to][]{Lyne2010}.
Small values of $C_1/C_2$ are associated with the mode clustered about a residual of zero (Mode A, shown in teal) while larger values are associated with the mode clustered around residuals of $\unit[-7.5]{\mu s}$ (Mode B, shown in orange).
The correlation between $C_1/C_2$ and the residuals confirms that the two modes are related to changes in $C_1/C_2$. We design a cut using $C_1/C_2$ that yields a subset of timing measurements targeting just a single mode\footnote{The ratio $C_0/C_2$ is also correlated with the post-fit residual, albeit less strongly than the ratio $C_1/C_2$.}.

\begin{figure}
	\centering 
	\includegraphics[width=\linewidth]{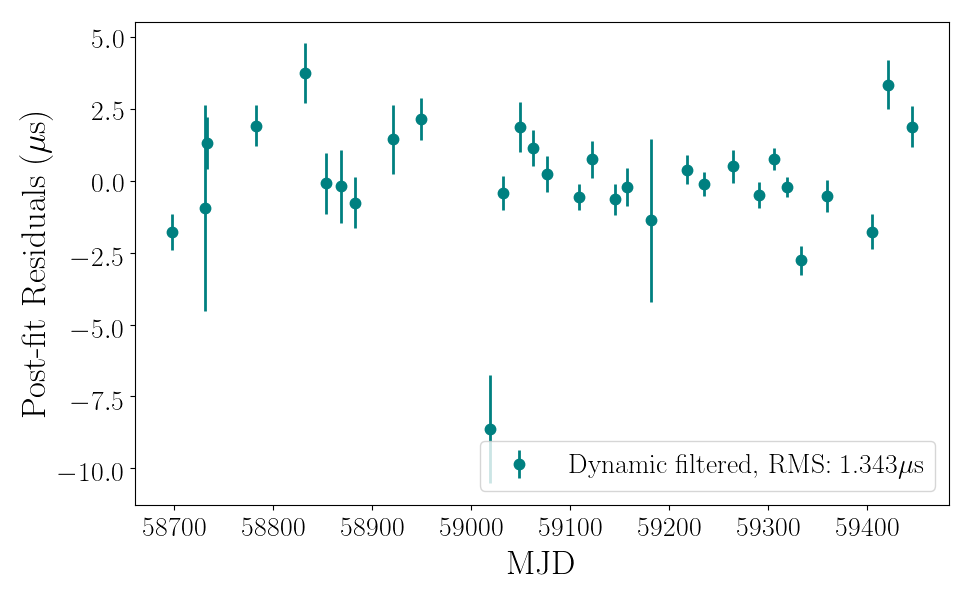}
	\caption{Timing solution produced with times of arrival from a flexible Gaussian fit, where pulses with a amplitude ratio ($C_1/C_2$) greater than 0.21 are removed. This timing solution has an RMS of $\unit[1.343]{\mu s}$, which is a factor of 1.78 improvement on the matched-filter template method. There is an outlier with a large negative residual at around 59020 MJD, due to imperfect separation between Mode A and Mode B when using the $C_1/C_2$ metric. }
	\label{fig:kookfilt}
\end{figure}

We cut the data, keeping epochs only if $C_1/C_2 < 0.21$ (the dashed vertical line in Fig.~\ref{fig:modechange}).
This cut results in the removal of the majority of Mode B while preserving the bulk of the Mode A epochs.
The cut is not perfect: there is still some contamination from Mode B epochs, and some of the data that is removed contains Mode A epochs.
This implies there are additional pulse-shape changes connected to the assumed mode-changing that this method is unable to capture.
However, the cut produces a more pure sample of Mode A.
Figure \ref{fig:kookfilt} shows the result of timing after the application of the cut.
Our timing solution has a root mean square error of only $\unit[1.343]{\mu s}$, which improves on the matched-filter template timing solution when all data is used, by a factor of 1.78. 

Next we determine how our analysis affects the usefulness of this data for searches for gravitational waves.
When modelling pulsar data, an uncorrelated white noise term is included to account for pulse jitter, the random variation in pulse shape\footnote{In pulsar timing, white noise is typically described by the sum of ECORR, EFAC, EQUAD \citep[see][for details]{IPTAVerbiest2016}.
When white noise is mentioned in this paper we refer only to ECORR, the pulse jitter term.}. This white-noise term in the full data set is $\unit[3.981]{\mu s}$, whereas in the data set with the cut it is $\unit[0.933]{\mu s}$. 
This factor 4.3 reduction in white noise suggests that the mode-changing behaviour was misspecified by the model, and was included instead as jitter noise. By accounting for mode-changing in the timing and reducing the white noise, our noise modelling can be more sensitive to time-correlated red noise processes, such as the stochastic gravitational-wave background. 

 The signal-to-noise ratio (SNR) of the gravitational-wave background scales as~\citep{Siemens_2013}
\begin{equation}
    \label{SNRGWB}
    \langle\rho\rangle \propto \left(c{\sigma^{-2}}\right)^{1/(2\beta)} ,
\end{equation}
where $\langle\rho\rangle$ is the (expectation value of the) signal-to-noise ratio of the gravitational-wave background, $\sigma$ is the white-noise RMS, $\beta$ is the spectral index of the gravitational-wave background, and $c$ is the observing cadence\footnote{Time correlated noise terms were not able to be modelled in this data set due to an insufficient time span, so instead of providing a direct upper limit for gravitational waves we provide a projection based on \citet{Siemens_2013}.}.
We use this scaling to assess the change in SNR from our analysis, which produces a smaller value for $\sigma$, but at the cost of throwing out some data:
\begin{align}
    \label{SNRGWBratio}
    \frac{ \langle\rho_\textrm{filtered}\rangle}{\langle\rho_\textrm{full}\rangle} = & \left(\frac{c_\textrm{filtered}\sigma_\textrm{filtered}^{-2}}{c_\textrm{full}\sigma_\textrm{full}^{-2}}\right)^{3/26}
    =1.32.
\end{align}
Thus, for this particular pulsar, we can achieve a 32\% reduction in the noise for a stochastic background analysis.

\section{Discussion}\label{discussion} 
We show, using PSR J1103$-$5403 as a case study, that employing a fitting method that can identify and then time a single mode (using flexible, epoch-to-epoch fitting) can improve timing solution residuals in mode-changing pulsars. 
This method may be able to identify other pulsars exhibiting mode changing and improve their timing. 
We hope to apply this method to other pulsars to see if we are able to reduce the inferred white noise similarly. 
Additionally, we hope to fit basis functions to pulsars experiencing other kinds of pulse-shape variability to see if these can be characterised and improved as well. Mode-changing was recently found in J1909-3744, which is the most precisely timed pulsar \citep{Miles2022Mode}. 
Therefore, dynamic pulse fitting may offer new insights to some of our most important and precise pulsars for timing. 

While this work primarily focuses on the improvement of timing precision for gravitational-wave detection, pulse-profile variability has the potential to provide insights into pulsar physics. 
In particular, the pulse emission mechanism is currently poorly understood, but the geometry of the pulsar magnetic field and emission heights are thought to influence the intensity of the observed radio flux. By using basis functions to characterise each observing epoch (or in the case of bright pulsars, individual pulses) we can start to piece together a picture of pulse-shape evolution. Large shape-change events such as mode changing, nulling and sub-pulse drifting have been studied in order to constrain pulsar emission models \citep[e.g.,][]{Kramer2006Mode, Lyne2010, Basu_emmision, Brook2016Mode, Brook2018Mode, Chen_emission, Shang_emission, Shaw2022Mode}, but we believe a broader data set of subtle, pulse-to-pulse, or longer time-span changes, may shed light on the emission mechanism further. Applying our method to many pulsars may therefore provide a large data set of pulse-shape evolution across multiple pulsars.

Mode-changing and nulling are related to interesting shape-change phenomena. They are both thought to arise from a change in the structure in the magnetosphere in the neutron star. \cite{Janagal2022shapechange} show a relationship between periods of nulling and mode changing in PSR J1822$-$2256, where nulling is always followed by a particular mode. This correlation suggests that the mechanisms between nulling and mode-changing may be strongly related. Mode changing and nulling are not commonly observed features of millisecond pulsars, but with PSR J1103$-$5403 being the fourth millisecond pulsar now observed to exhibit mode changing, it is clear that these events happen to pulsars with millisecond periods. This demonstrates that mode changing can occur on millisecond timescales, and this can continue to constrain the physical processes causing these phenomena. The observing epochs for PSR J1103$-$5403 are separated by days, but \cite{Miles2022Mode} show that the millisecond pulsar PSR J1909$-$3744 has an upper limit of \unit[2.941]{ms} on the time scale for magnetospheric changes causing the mode changes. 

There are a number of natural extensions to be made to the method introduced in this work. The observing epochs fit here are frequency- and polarisation-averaged. Pulse profiles are known to have a large dependence on frequency in many pulsars, and averaging over large bandwidths may induce larger timing solution residuals \citep{Demorest2013, Shannon2017ISM}. Low-frequency pulse-profile changes specifically offer insight about pulsar emission geometry, offering a rare study of the configuration of the pulse emission mechanism \citep{Olszanski2022}. Applying this method to frequency sub-banded data may yield a simple yet impactful extension. Additionally, polarisation profiles provide information about the electron density in the interstellar medium and magnetic fields in globular clusters \citep{Abbate2022}. The science to be gleaned from frequency and polarisation banded pulse profile fits motivates the application of Bayesian basis function fitting to them \citep[see][]{vanStraten2006, Lentati2017}. Given that each profile can be analyzed in parallel, this extra computation may be practical with a computing cluster, while offering the potential for rich pulsar science.

As discussed in Section \ref{fitting}, the choice for a three Gaussian model was based on a visual inspection of the average pulse morphology. This model worked well for PSR J1103$-$5403, but every pulsar will have a unique pulse shape and pulse-shape evolution. For this reason, the extension of this method to other pulsars may require different combinations of basis functions for each pulsar. The options are endless, but there is still the question of how many basis functions  are needed to best model data. To assist this, we are developing a trans-dimensional sampler that is able to fit the number of basis functions. This will allow us to remove the arbitrary choice of the number of basis functions.

Another pulsar of interest is PSR J1713$+$0747. This pulsar is observed by all constituents of the International Pulsar Timing Array due to its apparent stability and large signal-to-noise ratio. However, in April 2021 it underwent a dramatic pulse shape change event \citep{Lam_J1713, Singha1713, Jennings1713}, which induced large residuals in the timing solution using the matched-filter template approach. This shape change event seems to mainly effect the leading and trailing edges of the pulse, rather than the tallest component of the pulse. An appropriately modelled basis function fitting method would be able to reduce the timing residuals connected to these shape changes. In addition, quantifying the pulse-shape parameters of this event by recording how the parameters of the basis function change from pulse to pulse may offer insight as to how portions of the emission region are changing over the recovery period.

\section*{Acknowledgements}
We acknowledge and pay respects to the Elders and Traditional Owners of the land on which our institutions stand, the Bunurong and Wurundjeri Peoples of the Kulin Nation. The authors wish to thank Marcus Lower for his constructive feedback.
We wish to thank the anonymous reviewer for their helpful comments.
The authors acknowledge support from the Australian Research Council (ARC) Centre of Excellence CE170100004, Future Fellowship FT190100155, Discovery Projects DP220101610 and DP230103088. 
This work was performed on the OzSTAR national facility at Swinburne University of Technology. The OzSTAR program receives funding in part from the Astronomy National Collaborative Research Infrastructure Strategy (NCRIS) allocation provided by the Australian Government.
The MeerKAT telescope is operated by the South African Radio Astronomy Observatory, which is a facility of the National Research Foundation, an agency of the Department of Science and Innovation.

\section*{Data Availability}
The data underlying this article are available in the MeerKAT Pulsar Timing Array (MPTA) First Data Release on the gSTAR Data Sharing Portal, at \url{https://dx.doi.org/10.26185/6392814b27073}.

\bibliographystyle{mnras}
\bibliography{bib}

\appendix

\section{Priors on dynamic model parameters}
\label{AppendixA}
In Tab.~\ref{tab:priors} we show the prior distributions of the parameters fit to every observing epoch.

\begin{center}
    \begin{table}
        \begin{tabular}{ |c|c| } 
        \hline
         Parameter & Prior\\ 
         \hline
         $\phi_i$ (mean) & $\mathcal{U}\left(0,C_{i+1,i\neq2}\right)$ \\ 
         $\phi_2$ (mean) & $\mathcal{U}(0,1)$ \\ 
         $C_i$ (amplitude) & $\mathcal{U}\left(0,C_{i+1,i\neq2}\right)$ \\ 
         $C_2$ (amplitude) & $\mathcal{U}(0,1)$ \\ 
         $\beta_i$ (width) & $\log_{10}\left[\mathcal{U}\left(\frac{1}{1024}\times5,0.032\right)\right]$ \\ 
         \hline
        \end{tabular}
    \caption{The priors on the parameters fit to each epoch. There are three Gaussians for each fit, thus $i=\{0,1,2\}$. $\mathcal{U}(a,b)$ denotes a uniform prior distribution between $a$ and $b$.}
    \label{tab:priors}   

    \end{table}
\end{center}

\section{Posteriors of individual epoch fit}
\label{AppendixB}
\begin{figure*}
	\centering 
	\includegraphics[width=\linewidth]{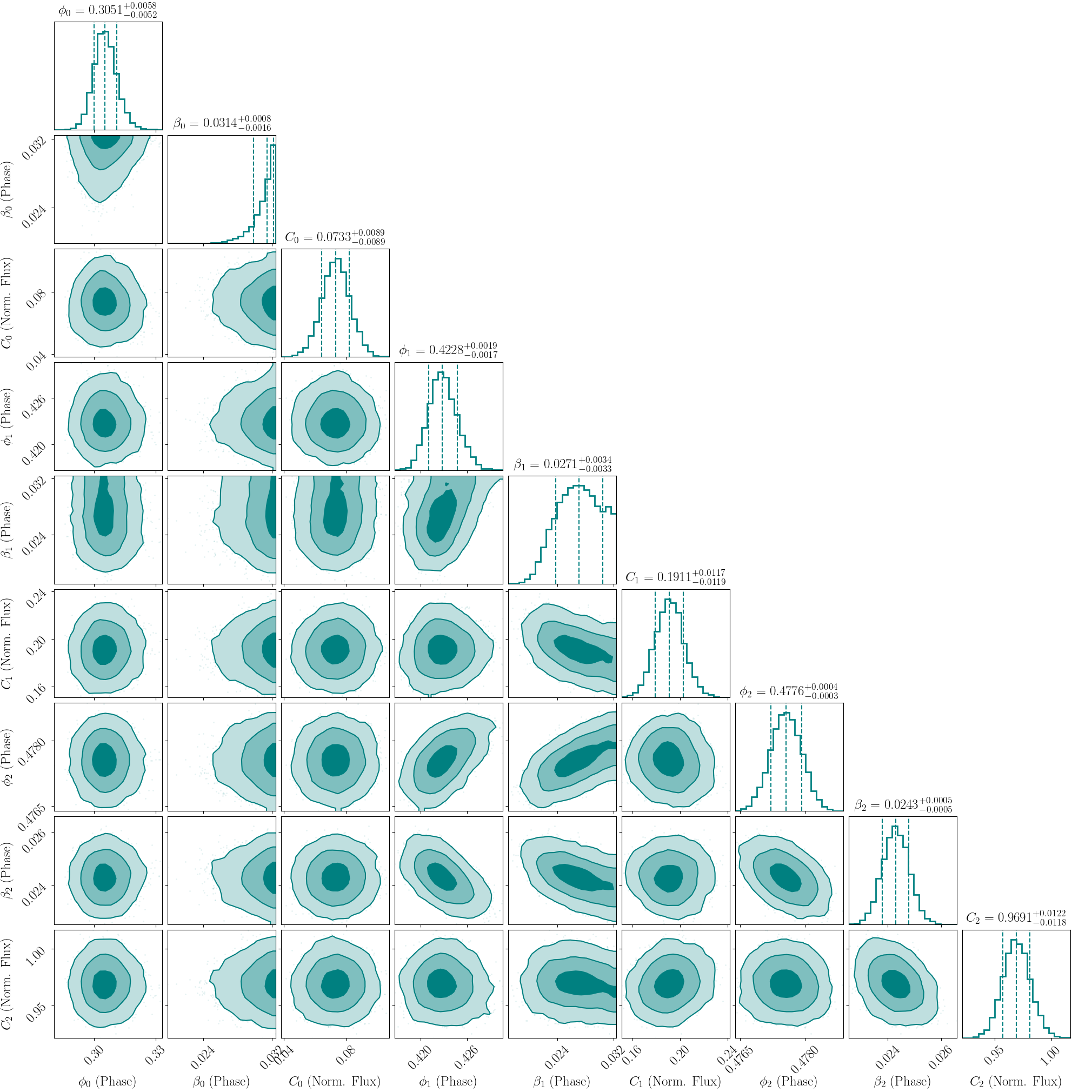}
	\caption{Posterior distribution of the parameters fit to the epoch shown in Fig.~\ref{fig:examplepulsefit}.  
	}
	\label{fig:corner}
\end{figure*}

In Fig.~\ref{fig:corner} we show the posterior distributions of the parameters of the three Gaussians fit to the epoch shown in Fig.~\ref{fig:examplepulsefit}. The $\phi_i$ represent the centering of the Gaussians in phase, the $\beta_i$ are the widths in phase and the $C_i$ are the amplitudes of each Gaussian in normalised flux. We use $\phi_2$ as the time of arrival for producing a timing solution. We place an arbitrary maximum on  $\beta_i$ to prevent the Gaussian fitting the shoulder of the pulse (Gaussian 1) from interfering with the main pulse (which we use to determine the time of arrival as described above). We note that our posterior has support at this arbitrary maximum (see Fig.~\ref{fig:corner}). However, we do not use the full posterior and, if we increase this maximum, the post-fit residuals are increased, therefore we argue that the arbitrary maximum is justified.

\end{document}